\documentstyle[preprint,prd,aps,psfig]{revtex}
\draft
\newcommand{\f}[2]{{#1 \over #2}}
\newcommand{\cZ}{{\cal Z}}
\newcommand{\Ref}[1]{(\ref{#1})}
\newcommand{\be}{\begin{equation}}
\newcommand{\ee}{\end{equation}}
\newcommand{\bn}{\begin{eqnarray}}
\newcommand{\en}{\end{eqnarray}}
\newcommand{\bd}{\begin{displaymath}}
\newcommand{\ed}{\end{displaymath}}
\newcommand{\bnn}{\begin{eqnarray*}}
\newcommand{\enn}{\end{eqnarray*}}
\begin{document}
\title{Ground State Energy of Massive Scalar Field Inside a Spherical Region 
in the Global Monopole Background}
\author{ E.R. Bezerra de Mello \thanks{e-mail:
emello@fisica.ufpb.br}, V.B. Bezerra \thanks{e-mail:
valdir@fisica.ufpb.br}, and N.R. Khusnutdinov \thanks{On leave from
Kazan State Pedagogical University, Kazan, Russia; e-mail:
nail@dtp.ksu.ras.ru}}
\address{ Departamento de F\'{\i}sica, Universidade Federal da
Para\'{\i}ba, \\ 
Caixa Postal 5008, CEP 58051-970 Jo\~ao Pessoa, Pb, Brazil}
\maketitle
\begin{abstract}
Using the zeta function regularization method we calculate the ground state 
energy of scalar massive field inside a spherical region in  the space-time 
of a point-like global monopole. Two cases are investigated: (i) We 
calculate the Casimir energy inside a sphere of radius $R$ and make an  
analytical analysis about it. We observe that this energy may be positive 
or negative depending on metric coefficient $\alpha$ and non-conformal 
coupling $\xi$. In the limit $R\to\infty$ we found a zero result. (ii) 
In the second model we surround the monopole by additional sphere of radius 
$r_0<R$ and consider scalar field confined in the region  between these two 
spheres.  In the latter, the ground state energy presents an  additional 
contribution due to boundary at $r_0$ which is divergent for small radius. 
Additional comments about renormalization are considered.  
\end{abstract}
\pacs{98.80.Cq, 14.80.Hv}
\section{Introduction}
Different types of topological objects may have been formed
during Universe expansion, these include domain walls, cosmic
strings and monopoles \cite{Kibble}.  These topological defects
appear as a consequence of breakdown of local or global gauge 
symmetries of a system composed by self-coupling iso-scalar Higgs
fields $\Phi^a$. Global monopoles are created due to phase transition 
when a global gauge symmetry is spontaneuosly broken  
and they may have been important for cosmology and astrophysics. 
The process of global monopole creation is accompanied by particles 
production \cite{Lousto}. Grand Unified Theory predicts great number of 
these objects in the Universe \cite{Preskill} but this problem may be 
avoided using inflationary models. From astrophysical point of view
there is at most one global monopole in the local group of
galaxies \cite{HiscockPRL}.  

The space-time of a global monopole in a $O(3)$ broken symmetry model
has been investigated by Barriola and Vilenkin
\cite{BarriolaVilenkin}. They have shown that far from the
compact monopole's core the space-time is approximately described
by spherical symmetric metric with an additional solid angle
deficit. It is also possible to find solution for the Einstein equation
coupled with an energy-momentum tensor associated with a pointlike global
monopole. For simplicity we shall consider in this paper this singular
configuration. In Ref. \cite{HarariLousto} a simplified model is presented
in order to consider some internal structure for the global monopole.  

The analysis of quantum fields on the global monopole background
have been considered in Refs. \cite{FieldsOnGlobalMonopole} - \cite{BBK}. 
It was shown, taking into account only dimensional and conformal considerations
\cite{FieldsOnGlobalMonopole}, that the vacuum expectation value of
the energy-momentum tensor associated with a collection of conformal massless 
quantum field of arbitrary spin in this background has the following 
general structure 
\bd
\langle T^i_k \rangle = S^i_k{\hbar c\over r^4}\ ,
\ed
where the quantities $S^i_k$ depend only on the solid angle deficit and spin
of the fields. For scalar field this tensor was investigated more carefully 
by Mazzitelli and Lousto \cite{Mazzitelli} and for massless 
spinor field by authors \cite{BBK} in great details.  

The above energy-momentum tensor has non-integrable singularity at
origin therefore the ground state energy cannot be found by
integrating the energy density. The same problem also appears for
cosmic string space-time \cite{FieldsOnCosmicString} and in
Minkowsky one with boundary condition on dihedral angle
\cite{Dihedrial}.  The calculation of ground state energy for the cosmic 
string space-time  was considered in Refs. \cite{Dima,KhusnutdinovBordag} 
using different approaches. For infinitely thin cosmic string, specific global 
effect appears which leads to additional surface renormalization \cite{Dima}.  
The ground state energy of massive scalar field in the background of 
a cosmic string with internal nonsingular structure has been considered in 
Ref. \cite{KhusnutdinovBordag}. It has been found that it is zero for 
arbitrary transverse diameter of the string. 

The nontrivial topological structure of space-time leads to a number of
interesting effects which are not presented in a flat space. For
example, there appear self-interacting forces on a massive point-like 
particles at rest. These forces have been investigated in 
Refs. \cite{Linet,Bezerra} for cosmic string and global monopole space-times 
respectively.

In the framework of zeta function regularization method 
\cite{ZetaFunction} (see also \cite{ElizaldeBook}) the
ground state energy of scalar massive field can be obtained by
\be
E(s) = {1\over 2}M^{2s}\zeta_A(s-{1\over 2})\ , 
\label{GroundEnergy}
\ee 
which is expressed in terms of the zeta function $\zeta_A $ associated with  
the Laplace operator $\hat A = -\triangle + \xi {\cal R} +m^2$ defined in the 
three  dimensional spatial section of the space-time. Here, the parameter 
$M$, with dimension of mass, has been introduced in order to give the correct 
dimension for the energy. In order to calculate the renormalized ground 
state energy we shall use the approach which was suggested and developed in 
Refs.
\cite{Method,Method2,Method3,mtoinfty}. 

In this paper we would like to discuss the ground state
energy of scalar massive field in the background of point-like global 
monopole space-time inside a spherical region considering an arbitrary
non-minimal coupling of this field with the geometry.
Because the energy-momentum tensor has non-integrable singularity at 
the origin we would like to investigate two cases: (i) In the first we 
consider  a point-like global monopole and calculate ground state energy using 
zeta function approach; (ii) in the second one, we consider a sphere 
surrounding the monopole and cut out internal part of it by an appropriate 
boundary condition for radial functions. This procedure permits us to reveal 
the role of the singularity. In the limit of zero radius of inner sphere, this 
model corresponds to topological defects because there is no internal 
structure of monopole for arbitrary radius of sphere.  

The zeta function of Laplace operator on the point-like global monopole 
background has been considered by Bordag, Kirsten and Dowker
in Ref. \cite{BordagKirstenDowker} using the method given in
Refs. \cite{Method,Method2,Method3,mtoinfty}. There, the general mathematical
structure of zeta function and the heat kernel coefficients on
the generalized cone have been  obtained. Because the main emphasis of the
present paper is on the ground state energy, we shall rederive in 
Sec.\ref{Zeta} some specific formulas for our case which was not considered 
in Ref. \cite{BordagKirstenDowker}.  

The organization of this paper is as follows. In Sec.\ref{Geometry} we briefly
review some geometrical properties about global 
monopole space-time which will be needed. In Sec.\ref{Zeta}, the zeta
function of the Laplace operator on three dimensional section of
a point-like global monopole space-time is developed. 
In Sec.\ref{Zeta1} we consider 
the zeta function for global monopole space-time cutting out by the 
sphere around the origin. In Sec.\ref{sec:GSE}, the ground state energy of 
massive scalar field with arbitrary non-conformal coupling on global monopole 
background is considered for both above cases. In Sec.\ref{last}, 
we discuss our results. The signature of the space-time, the sign of 
Riemann and Ricci tensors are the same as in Christensen paper 
\cite{Christensen}. We use units $\hbar = c= G = 1$. 


\section{The Geometry}\label{Geometry}
Global monopoles are heavy objects probably formed in the early
Universe by the phase transition which occur in a system
composed by a scalar self - coupling triplet field  $\phi^a$
whose original global symmetry $O(3)$ is spontaneously
broken to $U(1)$. 

The simplest model which gives rise a global monopole is
described by the Lagrangian density below
\bd
L= {1\over 2}(\partial_l \phi^a) (\partial^l \phi^a) - {\lambda
\over 4} (\phi^a \phi^a - \eta^2)^2\ .
\ed
Coupling this matter field with the Einstein equation, Barriola
and Vilenkin \cite{BarriolaVilenkin} have shown that the effect
produced by this object in the geometry can be approximately
represented by a solid angle deficit in the $(3+1)$ -
dimensional space-time, whose line element is given by
\be
ds^2=-dt^2 +\alpha^{-2}dr^2 + r^2(d\theta^2 + \sin^2\theta
d\varphi^2)\ ,\label{Metrica}
\ee
where the parameter $\alpha^2 = 1 - 8\pi\eta^2$ is smaller than
unity and depend on the symmetry breaking energy scale
$\eta$. The solid angle in the geommetry defined by \Ref{Metrica} is
$4\pi\alpha^2$, consequently smaller than $4\pi$. So this spacetime 
presents a solid angle defict given by $\delta\Omega=32\pi^2\eta^2$. We also
can note that it is not flat. The nonzero
components of Riemann and Ricci tensors, and scalar curvature are given below 
\bd
{\cal R}^{\theta\varphi}_{.\ .\theta\varphi}={\cal
R}^{\theta}_{\theta}= {\cal R}^{\varphi}_{\varphi}={1-\alpha^2
\over r^2}\ ,\ {\cal R}={2(1-\alpha^2)\over r^2}\ .
\ed
For further application let us consider extrinsic curvature tensor on
the sphere of radius $R$ around the origin
\bd
K_{ij}=\nabla_i N_j\ .
\ed
Here $N_j$ is outward unit normal vector with coordinates
$N_j=(0,\alpha,0,0)$. This tensor has two nonzero components
\bd
K^{\theta}_{\theta} = K^{\varphi}_{\varphi}={\alpha \over R}\ . 
\ed


\section{Zeta function for point-like global monopole space-time}\label{Zeta}
In order to calculate the ground state energy given by Eq.\Ref{GroundEnergy} 
we have to obtain the zeta function of the operator $\hat A$ in the 
neighborhood of point $s=-1/2$. For the calculation of zeta function we follow 
Refs \cite{Method2,Method3,BordagKirstenDowker}. 
The zeta function of the operator $\hat A = -\triangle + \xi {\cal R}
+ m^2$ is defined in terms of the sum over all eigenvalues of this
operator by 
\bd
\zeta_A(s-{1\over 2}) = \sum_{(n)}(\lambda_{(n)}^2 +
m^2)^{1/2-s}\ . 
\ed
Here $\lambda_{(n)}^2$ is the eigenvalue of operator 
$\hat B = \hat A - m^2$. The eigenfunctions of the operator $\hat A$ defined
in \Ref{Metrica} which are regular at the origin have the form 
\be
\Phi ({\bf r}) = \sqrt{\lambda \over \alpha r} Y_{lm}(\theta,
\varphi) J_\mu ({\lambda \over \alpha}r)\ , \label{Phi1}
\ee
where $Y_{lm}$ are the spherical harmonics and $J_\mu$ is the Bessel
function of the first kind with index 
\be
\mu = {1\over \alpha}\sqrt{(l+{1\over 2})^2 + 2(1 - \alpha^2)(\xi -
{1\over 8})}\ . 
\label{index}
\ee
A discret set of eigenvalues $\lambda_{l,j}$ can be found applying some 
boundary condition imposed on this function. Let us consider the
Dirichlet boundary condition at the surface of a sphere of
radius $R$ concentric with the pointlike monopole
\be
\sqrt{\lambda_{l,j}} J_\mu({\lambda_{l,j} \over \alpha}R)=0\ .
\label{Boundary} 
\ee 
Then, the zeta function reads 
\bd
\zeta_A^R(s-{1\over 2}) = \sum_{l=0}^{\infty} \sum_{j=0}^{\infty}
(2l+1)(\lambda_{l,j}^2+m^2)^{1/2 -s}\ ,
\ed
where the label $R$ in the zeta function was introduced to indicate 
this kind of boundary condition. The solutions $\lambda_{l,j}$ of 
equation \Ref{Boundary} can
not be found in closed form. For this reason we use the method
suggested in Refs. \cite{Method,Method2,Method3} which allows us to
express the zeta function in terms of the eigenfunctions. According
to this approach, the sum over $j$ may be converted into contour
integral in complex $\lambda$-plane using the principal of
argument, namely 
\bd
\zeta_A^R(s-{1\over 2}) = \sum_{l=0}^\infty (2l + 1) \int_\gamma
d\lambda^2 (\lambda^2 + m^2)^{1/2 - s}{\partial \over \partial
\lambda} \ln \lambda^{-\mu} J_\mu ({\lambda \over
\alpha} R)\ ,
\ed
where the contour $\gamma$ runs counterclockwise and must
enclose all solutions of Eq.\Ref{Boundary} on positive real
axis. Shifiting the contour to the imaginary axis we obtain
the following formula for the zeta function (see
\cite{Method2} for details) 
\be
\zeta_A^R(s-{1\over 2}) = -{\cos \pi s\over \pi}
\sum_{l=0}^\infty (2l + 1) \int_m^\infty dk (k^2 - m^2)^{1/2 -
s}{\partial \over \partial k} \ln k^{-\mu} I_\mu ({k \over
\alpha} R)\ .
\label{Main} 
\ee
Here $I_\mu$ is the modified Bessel function. Let us use the
uniform expansion for the Bessel function $I_\mu(\mu z)$ as
below 
\be
I_{\mu}(\mu z) = \sqrt{\frac{t}{2\pi \mu}} e^{\mu \eta (z)}
\left\{ 1 + \sum_{k=1}^\infty \frac{u_k(t)}{\mu^k} \right\}\ ,  
\label{UniformExpansion}
\ee
where $t = 1/\sqrt{1 + z^2}\ ,\ \eta (z) = \sqrt{1 + z^2} + \ln
(z/(1 + \sqrt{1 + z^2}))$ and $z = kR/\mu\alpha$. The firsts  
coefficients $u_k(t)$ and the recursion relations for higher ones are
listed in \cite{Abramowitz}. This uniform expansion leads to power 
series over $m$, and the term $u_N$ gives the contribution $\sim
1/m^{3-N}$. We shall make the calculations up to $N=3$. In this case we 
obtain the following formula for uniform expansion of the logarithm of Bessel 
function
\be
\ln(k^{-\mu}I_{\mu}({k \over \alpha}R))= \mu(\eta (z)- z) - {1\over 4} 
\ln (1 + z^2) + {1\over\mu}D_1(t) + {1\over\mu^2}D_2(t) + 
{1\over\mu^3}D_3(t)\ ,  \label{unex} 
\ee 
where 
\bn
D_1(t)&=& \sum_{a=0}^1x_{1,a}t^{1+2a}={1\over 8}t - 
{5\over 24}t^3\ , \nonumber \\
D_2(t)&=& \sum_{a=0}^2x_{2,a}t^{2+2a}={1\over 16}t^2 - {3\over 8}t^4 + 
{5\over 16}t^6\ , \\ 
D_3(t)&=& \sum_{a=0}^3x_{3,a}t^{3+2a}={25\over 384}t^3 - {531\over 640}t^5 + 
{221\over 128}t^7 - {1105\over 1152}t^9\ . \nonumber 
\en
In the above expression we omit all constants which are not important for the
calculation of zeta function. Adding and subtracting uniform expansion 
\Ref{unex} in integrand of formula \Ref{Main} we may represent the zeta 
function in the form
\be
\zeta_A^R(s-{1\over 2}) = N^R(s) +  {m^{-2s}\over (4\pi )^{3/2} 
\Gamma (s-{1\over 2})}\sum_{k=-1}^3 A_k(s,R)\ , \label{z}
\ee 
where 
\bn
N^R(s)&=& - {\cos\pi s\over\pi R} \sum_{l=0}^\infty (2l+1)
\mu\alpha\int_{\beta /\mu
\alpha}^\infty dx \left\{x^2 - \left({\beta \over \mu\alpha} 
\right)^2\right\}^{1/2 -s} \\  
&\times& {\partial \over \partial x} \left\{\ln I_\mu (\mu x) - 
\mu \eta (x) + {1\over 4} \ln (1 + x^2) - {1\over\mu}D_1(t) - 
{1\over\mu^2}D_2(t) - {1\over\mu^3}D_3(t)\right\}, \nonumber \\
A_{-1}(s,R)&=& {4\pi^{3/2} m\beta \over \alpha}  \sum_{l=0}^\infty (2l+1) 
\left[ 
\frac{\Gamma (s - 1)}{\sqrt{\pi}} {}_2F_1
-\frac{\alpha\mu}{\beta }\Gamma (s -\frac{1}{2}) \right]\ ,\label{A(-1)} \\
A_{0}(s,R)&=& -2\pi^{3/2} m \cZ (0,s-{1\over 2})\ , \\
A_{1}(s,R)&=& -{\pi m\alpha \over \beta} \left[\cZ (0,s) - {10 \over 3} 
\cZ (2,s+1) \right]\ , \\
A_{2}(s,R)&=& -{\pi^{3/2} m\alpha^2 \over 2\beta^2}\left[\cZ 
(0,s+{1\over 2}) - 
6\cZ (2,s+{3\over 2})+{5 \over 2} \cZ (4,s+{5\over 2}) \right]\ , \\ 
A_{3}(s,R)&=& -{\pi m\alpha^3 \over 24\beta^3}\left[25\cZ (0,s+1) - 
{1062 \over 5}\cZ (2,s+2) +{884 \over 5} \cZ (4,s+3)\right.\nonumber \\
&-&\left.{1768 \over 63} \cZ (6,s+4) \right]\ . \label{A3}
\en
Here ${}_2F_1 = {}_2F_1 (-{1\over 2}, s -1; {1\over 2}; -\left({\mu
\alpha \over \beta})\right)^2$ is the hypergeometric function, 
$\beta = mR$ and 
\be
\cZ (p,s) = \Gamma (q) \sum_{l=0}^\infty {2l+1\over (1 + \alpha^2 
\mu^2/\beta^2)^s}\left(\frac{\alpha \mu}{\beta}\right)^p\ .
\label{Zet}
\ee  
The series in Eq.\Ref{A(-1)}  
\be
T(s)=\sum_{l=0}^\infty (2l+1) \left[ \frac{\Gamma (s -
1)}{\sqrt{\pi}} {}_2F_1 -\frac{\alpha\mu}{\beta }\Gamma (s
-\frac{1}{2}) \right]  \ ,
\label{F21}
\ee
can be expressed in terms of the same function given in Eq.
\Ref{Zet}. Indeed, one can use analytical continuation of the
hypergeometrical function \cite{Abramowitz}
\bnn
&& {}_2F_1\left(-\frac{1}{2}, s-1; \frac{1}{2};
-\left(\frac{\alpha\mu}{\beta}\right)^2 \right) 
     = 
\frac{\alpha\mu}{\beta} \frac{\Gamma (1/2) \Gamma (s -
1/2)}{\Gamma (s-1)} \\ 
&+&
\frac{\Gamma (1/2) \Gamma (1/2 - s)}{\Gamma (-1/2) \Gamma (3/2 -
s)} \left( 1 + \left(\frac{\alpha\mu}{\beta} \right)^2
\right)^{1-s} {}_2F_1\left(1, s-1; s + \frac{1}{2}; \frac{1}{1 +
\left(\frac{\alpha\mu}{\beta}\right)^2 }\right) \ . 
\enn
So, the first term in the rhs. of the above equation cancels the
second one, divergent term in the sum \Ref{F21} which is
due to term $k^{-\mu}$ in \Ref{Main} (see Ref. 
\cite{KhusnutdinovBordag}).  Now, one can use power series
expansion for the hypergeometric function because its argument
$1/(1 + (\alpha\mu/\beta)^2)$ is always smaller than unity, so
we get 
\be
T(s)={1\over 2 \sqrt{\pi}}\Gamma (s-1/2) \sum_{l=0}^\infty
{\cZ(0,n+s-1) \over \Gamma (n+s +1/2)}\ . 
\label{T(s)}
\ee
Therefore in order to calculate  of the zeta function we have to obtain 
an analytical continuation to the series $\cZ(p,q)$. In fact we may
consider only 
\be
\cZ(0,s) = \Gamma (q)\sum_{l=0}^\infty {2l+1\over (1 +
\alpha^2\mu^2/\beta^2)^s}\ ,
\label{Zet0}
\ee
because the other functions with $p=2,4,6,..$ can be expressed in
terms of $\cZ(0,q)$ only. Substituting the value for $\mu$ given in
Eq.\Ref{index} into Eq.\Ref{Zet0} we obtain 
\be
\cZ(0,s)=2\Gamma (q)\beta^{2s}\sum_{l=0}^\infty {l+1/2\over
((l+1/2)^2 + b^2)^s}\ ,
\ee
where $b^2 = \beta^2 + 2(1-\alpha^2)(\xi -1/8)$. This series
is convergent for $\Re q >1$. It is no difficult to obtain the analytical 
continuation of this series for small value of parameter $b$. Indeed, 
expanding $\cZ$ in powers of $b$ we have 
\be
\cZ(0,s)=2\beta^{2s}\sum_{k=0}^\infty {(-1)^k \over k!} \Gamma (k+s) 
b^{2k}\zeta_H(2k+2s-1,{1\over 2})\ . \label{bsmall}
\ee
For analytical continuation of this function in the domain $\Re q\leq 1$ and 
for great value of $b$, let us consider the series below 
\bd
F(s,a,b^2) = \sum_{l=0}^\infty {1 \over ((l+a)^2 + b^2)^s}\ . 
\ed
This series, which has been considered in great detail by Elizalde
\cite{Elizalde}, presents the following analytical continuation for 
great $b$ 
\bnn
F(s,a,b^2) &\simeq& {b^{-2s}\over \Gamma (s)} \sum_{l=0}^\infty {(-1)^l
\Gamma (l+s)\over l!} b^{-2l} \zeta_H(-2l,a) + {\sqrt{\pi}
\Gamma (s-1/2) \over 2\Gamma (s)}b^{1 -2s} \nonumber \\
&-& {2\pi b^{-1/2 - s}\over \Gamma (s)} \sum_{n=1}^\infty
n^{s-1/2}\cos (2\pi n a) K_{s-1/2} (2\pi n b)\ .
\enn
Here $\zeta_H$ is the Hurwitz zeta function and $K_n$ is the
modified Bessel function. Differentiating this series with
respect to $a$ and putting $a=1/2$ we obtain the analytical
continuation that we need, which is the following
\bd
\sum_{l=0}^\infty {l+1/2 \over ((l+1/2)^2 + b^2)^s} \simeq {b^{2-2s}
\over 2(s-1)} + \sum_{l=0}^\infty {(-1)^l \Gamma (l+s) \over l!
\Gamma (s)} b^{-2s - 2l}\zeta_H(-1 -2l,1/2)\ .
\ed
Taking into account this expression we obtain analytical
continuation for function $\cZ(0,q)$:
\be
\cZ(0,s)\simeq \left({b^2\over \beta^2}\right)^{-s} \left\{ b^2 \Gamma
(s-1) + 2 \sum_{l=0}^\infty {(-1)^l \over l!} \Gamma (l+s)
b^{-2l} \zeta_H(-1-2l,1/2) \right\}\ ,
\label{Zet0An}
\ee
where $b^2/\beta^2 = 1 + 2(1-\alpha^2)(\xi - 1/8)/\beta^2$. This
function has simple poles for integer numbers $q = 1,0,-1,-2,...$.
In order to obtain a renormalized value for the ground state energy 
we have to extract from our 
expression for zeta function \Ref{z} the part which survives in the limit 
$m\to\infty$.  Moreover to calculate the zeta function up to degree $m^0$ we
need only two terms from series \Ref{Zet0An} in which 
$\zeta_H(-1,1/2) =1/24\ ,\ \zeta_H(-3,1/2) = - 7/960$ and three
terms of $T(s)$  which are given by Eq.\Ref{T(s)}.  

Putting this expression into Eq.\Ref{T(s)}, and Eqs.\Ref{A(-1)} - \Ref{A3} and
expanding over $1/\beta = 1/mR \ll 1$ and $s$, also collecting
terms with similar degree on the mass $m$ up to $m^0$ (we cannot
here collect higher orders of $m$ because we used 
uniform expansion up to this power) we get 
\bn
\zeta_A^R (s-\f 12) &=& {m^{-2s} \over (4\pi )^{3/2}} \left\{
\left[{4\pi R^3 \over 3\alpha}\right]m^4 {\Gamma (s-2)\over
\Gamma (s-\f 12)} + \left[-2\pi^{3/2}R^2\right]m^3{\Gamma (s-\f 32)
\over \Gamma (s-\f 12)} \right. \label{DSR}\\
&+& \left[ \f 73 \pi \alpha R - {4\pi R
\over \alpha}(\Delta - \f 1{12}) \right]m^2 {\Gamma (s-1)
\over \Gamma (s-\f 12)} + \left[ 2\pi^{3/2} (\Delta - \f 1{12}) 
\right]m \nonumber \\
&+&\left. \left[{\pi\alpha \over R} (\Delta - \f 1{12}+\f{229}{2520}\alpha^2) 
- {2\pi \over \alpha R}
(\Delta^2 - \f 16\Delta + \f 7{240}) \right]
{\Gamma (s) \over \Gamma (s- \f 12)} + \dots \right\}\ . \nonumber
\en
Here $\Delta = 2(1-\alpha^2)(\xi -1/8)$. All these terms are poles
contributions in zeta function, all next terms will be finite
for $s\to 0$. Comparing the above expression with that obtained by
the Mellin transformation over trace of heat kernel (in three
dimensions) 
\bn
\zeta_A^R(s-1/2) &=& {1 \over \Gamma (s-\f 12)} \int_0^\infty dt
t^{s-3/2} K(t) = {m^{-2s} \over (4\pi )^{3/2}} \left\{ B_0^R m^4
{\Gamma (s-2) \over \Gamma (s-\f 12)} \right. \label{ze}\\
&+& \left. B_{\f 12}^R m^3 {\Gamma (s-\f 32)
\over \Gamma (s-\f 12)} + B_1^R m^2 {\Gamma (s-1) \over \Gamma
(s-\f 12)} + B_{\f 32}^R m + B_2^R {\Gamma (s) \over \Gamma (s-\f 12)}
 + \dots \right\}\ , \nonumber
\en
we obtain the heat kernel coefficients : 
\bn
B_0^{R} &=& {4\pi R^3 \over 3\alpha}\ ,\ B_{\f 12}^R= -2\pi^{3/2}R^2\ ,\
B_1^R = \f 73 \pi \alpha R - {4\pi R \over \alpha}(\Delta -
\f 1{12})\ ,\ \label{hkc} \\ 
B_{\f 32}^R&=& 2\pi^{3/2} (\Delta - \f 1{12})
\ ,\ B_2^R = {\pi\alpha \over R} (\Delta -\f 1{12}+\f {229}{2520}\alpha^2) 
- {2\pi \over \alpha R}
(\Delta^2 - \f 16\Delta + \f 7{240}) \ .\nonumber
\en
Those terms which are proportional to inverse degree of $\alpha$
come from exponential part of the uniform expansion given by 
\Ref{UniformExpansion}, and respectively for $T(s)$ \Ref{T(s)}.
The terms which are linear in $\alpha^1$ or $\alpha^0$ come from the
series $\sum u_k/\mu^k$ in \Ref{UniformExpansion}. 

Now we may compare our results with well-known formulas given in
Refs. 
\cite{KennedyCritchleyDowker,BordagKirstenDowker,ElizaldeBook,BransonGilkeyVass}. 
The coefficients $B_0^R,\ B_{\f 12}^R,\ B_1^R,\ B_{\f 32}^R$ coincide with 
general formulas in three dimensions (all geometrical quantities are given in
Sec.\ref{Geometry})
\bn
B_0^R &=&\f{4\pi R^3}{3\alpha}= \int_V dV\ , \label{B01/23/2} \\
B_{\f 12}^R &=& -2\pi^{3/2} R^2= -\f{\sqrt{\pi}}2 \int_{\partial
V} dS\ ,\ \nonumber \\
B_1^R &=& \f 73 \pi \alpha R - {4\pi R \over \alpha}(\Delta -
\f 1{12}) = \left(\f 16 - \xi\right)\int_V{\cal R}dV + \f 13 \int_{\partial V} 
\left(tr K\right)dS\ , \nonumber \\
B_{\f 32}^R &=& 2\pi^{3/2}\left(\triangle - \f 1{12}\right)\nonumber \\
&=& -\f{\sqrt{\pi}}{192} \int_{\partial V}\left( -96\xi{\cal
R} + 16 {\cal R} + 8 {\cal R}_{ik}N^iN^k + 7(trK)^2 - 10\left(trK\right)^2
\right) dS\ . \nonumber
\en 
However some problems are connected with the term $B_2$. The general structure
of this term is the following (see \cite{Christensen,BransonGilkeyVass}, 
for example) 
\be
B_2^R = \int_V b_2dV + \int_{\partial V}c_2dS\ ,\label{b2main}
\ee
where
\be
b_2 = -{1 \over 180} {\cal R}^{ik}{\cal R}_{ik} + {1 \over
180}{\cal R}^{iklj}{\cal R}_{iklj} + {1\over 6}\left({1\over 5}
- \xi\right)\Box {\cal R} + {1\over 2}\left({1 \over 6}
-\xi\right)^2{\cal R}^2 \ ,
\ee
is volume part, and 
\bn
c_2 &=&\f 13\left(\f 16 - \xi\right){\cal R}\left(trK\right)+
\f 13\left(\f 3{20}-\xi\right){\cal R}_{;l}N^l-\f 1{90}
{\cal R}_{lk}N^lN^k\left(trK\right) + \f 1{30}{\cal R}_{iljk}
N^lN^k K^{ij}\nonumber\\ 
&-& \f 1{90}{\cal R}_{il} K^{il} 
+\f 1{315}\left[\f 53 \left(tr K\right)^3 - 11\left(trK\right)\left(trK^2
\right) + \f{40}3\left(trK^3\right)\right] + \f 1{15}\Box \left(trK\right)\ ,
\en 
is boundary contribution. Taking into account the results obtained in
Sec.\ref{Geometry} we have these terms in  manifest form:
\bn
b_2&=&-\f 1{r^4}\f\alpha{4\pi}\left\{\pi\alpha\left(\triangle - \f 1{12} + 
\f{17}{120}\alpha^2\right) - \f{2\pi}{\alpha}\left(\triangle^2 - 
\f 16\triangle + \f 7{240}\right)\right\}\ ,\\
c_2&=&-\f{4\alpha^3}{315R^3}\ .
\en
We observe that the $b_2$ is proportional to $1/r^4$ and the integral over 
volume in Eq. \Ref{b2main} will diverge at origin. This problem has already
been discussed by Cheeger \cite{Cheeger}, Br\"uning and Seeley 
\cite{Bruning} and Bordag, Kirsten and Dowker \cite{BordagKirstenDowker} 
using {\em partie finite} of the integral. We regularize the 
expression for $B_2$ by restricting domain of radial integration  
\be
B_2=-\int_\varepsilon^R\f{dr}{r^2}\left\{\pi\alpha\left(\triangle - 
\f 1{12} + \f{17}{120}\alpha^2\right) - \f{2\pi}{\alpha}\left(\triangle^2 - 
\f 16\triangle + \f 7{240}\right)\right\}-\f{16\pi\alpha^3}{315R}\ .
\label{B2}
\ee
After integration we take its finite remainder parts as $\varepsilon\to 0$, 
and the expression obtained in this way coincides with that given in 
Eq.\Ref{hkc}. 

Our expressions for the heat kernel coefficients also agree with that ones
obtained in Ref.\cite{BordagKirstenDowker}. In that paper the heat kernel 
coefficients have been calculated for conformal case ($\xi = 1/8$ in three 
dimensions). In order to compare both results we have to set 
$\xi = 1/8\ (\triangle =0)$ in Eq.\Ref{hkc} and use the formulas of 
Appendix A from Ref.\cite{BordagKirstenDowker} for the three dimensional 
case $d=2$.

For renormalization which we shall discuss later we shall extract from 
zeta function \Ref{z} the asymptotic expansion \Ref{DSR}. Because all 
divergences at $s\to 0$ are contained in \Ref{DSR}, we set $s=0$
in the remained part. After long calculation we arrive at the following 
formula for zeta function 
\bn
\zeta_A^R(s-{1\over 2})&=& -{m\over 16\pi^2 \beta}\left\{B^R(\beta)\ln\beta^2  
+ \Omega^R (\beta)\right\}\label{ZetaR}+{m^{-2s} \over (4\pi )^{3/2}} \left\{ 
B_0^R m^4{\Gamma (s-2) \over \Gamma (s-{1\over 2})}\right.\nonumber \\ 
&+&\left. B_{\f 12}^R m^3 {\Gamma (s-{3\over 2})
\over \Gamma (s-{1\over 2})} + B_1^R m^2 {\Gamma (s-1) \over \Gamma
(s-{1\over 2})} + B_{\f 32}^R m + B_2^R {\Gamma (s)\over\Gamma (s-{1\over 2})} 
\right\}\ , 
\en
where 
\be
B^R(\beta) = {1\over 2}Rm^4B_0^R - Rm^2B_1^R + RB_2^R = 
{1\over 2} \beta^4b_0^R - \beta^2b_1^R +  b_2^R\ . \label{B^R}
\ee
In order to exhibit the dependence on mass the $m$ and on the radius of 
sphere $R$, we have introduced in above formula, the dimensionless heat kernel 
coefficients by relations 
\bn
b_0^R &=& B_0^R/R^3=\f{4\pi}{3\alpha}\ ,\  b_1^R = B_1^R/R=\f 73 
\pi \alpha  - \f{4\pi}\alpha(\Delta -\f 1{12})\ ,\\ 
b_2^R &=& B_2^R R=\pi\alpha (\Delta - \f 1{12} + \f{229}{2520}\alpha^2) - 
\f{2\pi}\alpha (\Delta^2 - \f 16\Delta + \f 7{240}) \ .
\en
The function $\Omega^R (\beta)$ tends to a constant for $\beta \to 0$ 
and $\Omega^R (\beta) = -B^R(\beta) \ln\beta^2 + \sqrt{\pi}b_{5/2}^R/\beta + 
O(1/\beta^2)$ for $\beta \to\infty$. The details of calculation and close 
form of $\Omega^R (\beta)$ are outlined in Appendix \ref{appA}. 

At this point we would like to make a comment. The origin of the term 
$B^R \ln\beta^2$ is the following: In the limit $m\to\infty$ the singular 
part of zeta function has the structure given 
by Eq.\Ref{ze}. For small value of $m$ it has the same poles structure 
multiplied by $\beta^{2s}$. This is because all functions $\cZ (p,s)$ 
are proportional to this degree of $\beta$ as it may be seen from 
Eq.\Ref{bsmall}. The difference between them in the limit 
$s\to 0$ is $s \ln\beta^2$ multiplied by Eq.\Ref{ze}. Obviously that in 
this limit only $B_0^R,\ B_1^R,\ B_2^R$ survive which give the 
logarithm contribution to Eq.\Ref{ZetaR}.


\section{The Model}\label{Zeta1}

Because the geometrical characteristics of global monopole space-time 
are divergent at the origin we consider the following model: The 
center of monopole is surrounded by sphere with radius $r_0$ whose 
interior region is cut out. It means that in our model there is no an 
internal structure for the  global monopole. The present model reflects  
this peculiarity of topological defect. 

In frameworks of this model, we have to take into account both solutions 
of radial equation of the Laplace operator, instead of only one given in 
Eq.\Ref{Phi1} which is regular at origin. The eigenfunctions now have the 
following form 
\be
\Phi ({\bf r}) = \sqrt{\lambda \over \alpha r} Y_{lm}(\theta,
\varphi)\left\{ C_1J_\mu ({\lambda \over \alpha}r) + 
C_2N_\mu ({\lambda \over \alpha}r)\right\}\ , \label{Phi2}
\ee  
where $N_\mu$ is the Bessel function of the second kind.

In this case we have two boundaries and one has to impose two boundary 
conditions. Let us again choose the Dirichlet boundary condition for the 
radial functions at spheres of radii $R$ and  $r_0$:
\be
C_1J_\mu ({\lambda \over \alpha}R) + C_2N_\mu ({\lambda \over \alpha}R) 
= 0\ ,
\ee
and
\be
C_1J_\mu ({\lambda \over \alpha}r_0) + C_2N_\mu ({\lambda \over \alpha}r_0)
=0\ .
\ee 
The set of discret eigenvalues $\lambda_{l,j}$ can be found from equation below
\be
J_\mu ({\lambda_{l,j} \over \alpha}r_0)N_\mu ({\lambda_{l,j} \over \alpha}R) - 
N_\mu ({\lambda_{l,j} \over \alpha}r_0)J_\mu ({\lambda_{l,j} \over \alpha}R) 
=0\ , \label{condition2}
\ee
which is, in fact, the condition for existence of the solution \Ref{Phi2}. 
Therefore we obtain the following formula for zeta function instead of 
Eq.\Ref{Main} 
\bn
\zeta_A(s-{1\over 2}) &=& -{\cos \pi s\over \pi}\sum_{l=0}^\infty (2l + 1) 
\int_m^\infty dk (k^2 - m^2)^{1/2 -s}\nonumber\\
&\times&{\partial \over \partial k}\ln\left(I_\mu ({kR\over\alpha})
K_\mu ({kr_0\over\alpha}) - K_\mu ({kR\over\alpha}) 
I_\mu ({kr_0\over\alpha})\right)\ .
\label{Main1} 
\en
This general expression may be essentially simplified in the limit 
$R/r_0\to\infty$ which we are interested in. Taking into account that 
in this limit  the ratio $K_\mu(kR/\alpha)/I_\mu(kR/\alpha) < \pi 
\exp (-2mR/\alpha)$ is exponentially small, so we may divide the expression 
for zeta function \Ref{Main1} in two parts
\be
\zeta_A(s-{1\over 2})=\zeta_A^R(s-{1\over 2})+\zeta_A^{r_0}(s-{1\over 2})\ ,
\label{division}
\ee 
where
\be 
\zeta_A^R(s-{1\over 2}) = -{\cos \pi s\over \pi}
\sum_{l=0}^\infty (2l + 1) \int_m^\infty dk (k^2 - m^2)^{1/2 -
s}{\partial\over\partial k}\ln\left(k^{-\mu}I_\mu ({k\over\alpha} R)\right)\ , 
\label{ZR}\ 
\ee
and
\be
\zeta_A^{r_0}(s-{1\over 2}) = -{\cos \pi s\over \pi}
\sum_{l=0}^\infty (2l + 1) \int_m^\infty dk (k^2 - m^2)^{1/2 -
s}{\partial \over \partial k}\ln\left(k^{\mu} K_\mu ({k\over\alpha} r_0) 
\right)\ . \label{Zr0}
\ee
The first part is the zeta function for pointlike global monopole which we 
have already calculated in last section. It depends only on the boundary 
condition on the sphere of radius $R$. The second part depends on 
boundary condition on the inner sphere of radius $r_0$. This kind of division 
of zeta function has been taken place for the case of thick cosmic string in 
Ref. \cite{KhusnutdinovBordag}. 
It is also in qualitative agreement with \cite{Method3}. Indeed, according 
with \cite{Method3}, the internal solution gives Bessel function $I_\mu$ and 
the external solution gives function $K_\mu$ in expression for zeta function. 
The first part of zeta function \Ref{ZR} depends on the 
solutions which are internal with respect of sphere of radius $R$ and the
second part of zeta function \Ref{Zr0} depends on the solutions which 
are external for sphere of radius $r_0$. 

Let us consider now the second expression \Ref{Zr0}. 
To calculate  $\zeta_A^{r_0}$ we use the same approach which we have 
used in last section. We have to take into account the uniform expansion for 
modified Bessel function of second kind $K_\mu (\mu x)$ which has the form 
below
\be
K_{\mu}(\mu z) = \sqrt{\frac{\pi t}{2\mu}} e^{-\mu \eta (z)}
\left\{ 1 + \sum_{k=1}^\infty (-1)^k\frac{u_k(t)}{\mu^k} \right\}\ .  
\label{UniformExpansionofK}
\ee
Differently from the uniform expansion of Bessel function of first kind given 
by Eq.\Ref{UniformExpansion}, the odd degrees of $\mu$ in above formula have 
the opposite sign. This fact leads to the change of sign of the heat kernel 
coefficients  with integer index, also with respect of the heat kernel 
coefficients which were considered in last section. Using this uniform 
expansion we arrive at the following formulas for the zeta function 
$\zeta_A^{r_0}$ 
\be
\zeta_A^{r_0}(s-{1\over 2}) = N^{r_0}(s) +  {m^{-2s}\over (4\pi )^{3/2} 
\Gamma (s-{1\over 2})}\sum_{k=-1}^3 (-1)^k A_k(s,r_0)\ , 
\ee
where   
\bn
N^{r_0}(s)&=& - {\cos\pi s\over\pi R} \sum_{l=0}^\infty (2l+1)\mu\alpha\int_{
\beta /\mu \alpha}^\infty dx \left\{x^2 - \left({\beta \over \mu\alpha} 
\right)^2\right\}^{1/2 -s}\\ 
&\times&{\partial \over \partial x} \left\{\ln(K_\mu 
(\mu x)) + \mu \eta (x) + {1\over 4} \ln (1 + x^2)
+{1\over\mu}D_1(t) - {1\over\mu^2}D_2(t) + 
{1\over\mu^3}D_3(t)\right\}\ , \nonumber 
\en
and the functions $A_k(s,r_0)$ are the same as in Eqs.\Ref{A(-1)} - \Ref{A3} 
but they depend now on the radius $r_0$. Proceeding in the same way as it 
was done in last section, we obtain the following expression for second 
part of zeta function $\zeta^{r_0}_A$: 
\bn
\zeta_A^{r_0}(s-\f 12)&=& -{m\over 16\pi^2 \beta_0}\left\{B^{r_0}(\beta_0)
\ln\beta_0^2  + \Omega^{r_0} (\beta_0)\right\}\label{Zetar0}+ 
{m^{-2s} \over (4\pi )^{3/2}} \left\{ B_0^{r_0} m^4
{\Gamma (s-2) \over \Gamma (s-\f 12)}\right.\\
&+&\left. B_{1/2}^{r_0} m^3 {\Gamma (s-\f 32)
\over \Gamma (s-\f 12)} + B_1^{r_0} m^2 {\Gamma (s-1) \over \Gamma
(s-\f 12)} + B_{3/2}^{r_0} m + B_2^{r_0} {\Gamma (s) \over \Gamma (s-\f 12)} 
\right\}\ , \nonumber
\en
where $\beta_0 = mr_0$ and 
\be
B^{r_0}(\beta_0) = \f 12 r_0m^4B_0^{r_0} - r_0m^2B_1^{r_0} + r_0B_2^{r_0} = 
\f 12 \beta_0^4b_0^{r_0} - \beta_0^2b_1^{r_0} +  b_2^{r_0}\ .
\ee
Contrary to the last section the heat kernel coefficients with integer number 
have changed the sign and they are
\bn
b_0^{r_0} &=& B_0^{r_0}/r_0^3=-\f{4\pi}{3\alpha}\ ,\  b_1^{r_0} = 
B_1^{r_0}/r_0=-\f 73\pi\alpha + {4\pi  \over \alpha}(\Delta -\f 1{12})\ ,\\ 
b_2^{r_0} &=& B_2^{r_0} r_0=-\pi\alpha (\Delta - \f 1{12} + 
\f{229}{2520}\alpha^2) + \f{2\pi}{\alpha}
(\Delta^2 - \f 16\Delta + \f 7{240}) \ .
\en
The summary heat kernel coefficients, according to Eq.\Ref{division}, are 
the sum of $B_n^R$ and $B_n^{r_0}$ and they are in agreement with general 
formulas. We have to take into account that normal vectors for sphere of 
radius $R$ and $r_0$ have opposite directions and that the boundaries consist 
now of two spheres. It is easy to understand the division of zeta function in 
two parts given by Eq.\Ref{division}, and opposite sign of the heat kernel 
coefficients $B^R$ and $B^{r_0}$ with integer indexes, by calculating $B_0$ 
and $B_{\f 12}$. For the space between two spheres we have 
\bn
B_0 &=& \int_V dV = \f{4\pi}\alpha \int_{r_0}^R r^2 dr=\f{4\pi}{3\alpha}R^3 - 
\f{4\pi}{3\alpha}r_0^3 = B_0^R + B_0^{r_0}\ , \\
B_{\f 12} &=& -\f{\sqrt{\pi}}2 \int_R dS  -\f{\sqrt{\pi}}2 \int_{r_0} dS = 
-2\pi^{3/2}R^2 -2\pi^{3/2}r_0^2   =  B_{\f 12}^R + B_{\f 12}^{r_0}\ .  
\en   
Therefore the full zeta function in this case has the following form
\bn
\zeta_A(s-\f 12)&=& -{m\over 16\pi^2 \beta_0}\left\{B^{r_0}(\beta_0)
\ln\beta_0^2  + \Omega^{r_0} (\beta_0)\right\} - {m\over 16\pi^2 \beta}
\left\{B^{R}(\beta)\ln\beta^2  + \Omega^R (\beta)\right\} 
\label{ZetaFull} \\ 
&+& 
{m^{-2s} \over (4\pi )^{3/2}} \left\{ B_0 m^4
{\Gamma (s-2) \over \Gamma (s-\f 12)}
+ B_{\f 12} m^3 {\Gamma (s-\f 32)
\over \Gamma (s-\f 12)} + B_1 m^2 {\Gamma (s-1) \over \Gamma
(s-\f 12)} + B_{\f 32} m \right.\nonumber\\
&+&\left. B_2 {\Gamma (s) \over \Gamma (s-\f 12)} \right\}\ . \nonumber
\en
The close expression for $\Omega^{r_0}$ is given in Appendix \ref{appA}. 


\section{The ground state energy}
\label{sec:GSE}

In the framework of zeta function approach the ground state energy is 
proportional to the zeta function of Laplace operator and is given by 
Eq.\Ref{GroundEnergy}. In order to analyze this energy, let us first of all, 
consider the ground state energy 
for point-like global monopole. The full energy of the system consists of 
two parts, namely classical part due to the boundary and monopole background, 
and quantum one loop correction. The general expression for boundary 
contributions has been considered in Refs \cite{ZetaFunction,Method3} and it 
has the following form:
\be
E_{cl}^R=p_RV_R +\sigma_R S_R + F_RR + \Lambda_R + \f {h_R}R\ . \label{Ecl}
\ee
Here $V_R=4\pi R^3/3\alpha$ and $S_R=4\pi R^2$ are the volume and area of 
spherical surface, respectively. The two parameters $p_R$ and $\sigma_R$ have 
simple physical means as pressure and tension of surface. The constant 
contribution described by parameter $\Lambda_R$ may be explained by the 
cosmological constant \cite{Brown}.  The other two parameters 
$F_R,\ h_R$ have not got a special names. 

The energy of monopole background can be obtained by integrating the $(t,t)$  
component of energy momentum tensor \cite{FieldsOnGlobalMonopole} 
\be
E_{cl}^{gm} =-\int_0^R \f{\eta^2\alpha^2}{r^2}dV=-4\pi\eta^2\alpha R\ .
\ee 
The quantum correction, using Eq.\Ref{ZetaR}, is 
\bn
E_q^R &=& \f 12 M^{2s}\zeta_A^R(s-{1\over 2})_{s\to 0}
= -{m\over 32\pi^2 \beta}\left\{B^R(\beta)\ln\beta^2  
+ \Omega^R (\beta)\right\}\label{QCR}+\left(\f Mm\right)^{2s} 
{1\over 16\pi^{3/2}} \nonumber \\
&\times&\left\{ B_0^R m^4
{\Gamma (s-2) \over \Gamma (s-{1\over 2})}
-\f 23 B_{\f 12}^R m^3 + B_1^R m^2 {\Gamma (s-1) \over \Gamma
(s-{1\over 2})} + B_{\f 32}^R m + B_2^R {\Gamma (s)\over\Gamma (s-{1\over 2})} 
\right\}_{s\to 0}\ , 
\en
where the heat kernel coefficients $B_k^R$ and $B^R$ are given by Eq.\Ref{hkc} 
and \Ref{B^R}, respectively. 

In order to obtain a well defined result for the full energy, we have to 
renormalize the parameters of classical part \Ref{Ecl} according to the rules 
below:
\bn
p_R &\to& p_R - \left(\f Mm\right)^{2s} {3 m^4b_0^R\over 64\pi^{5/2}}
{\Gamma (s-2) \over \Gamma (s-{1\over 2})}\ ,\ 
\sigma_R\to\sigma_R +{m^3b_{1/2}^R\over 96\pi^{5/2}}\ ,\nonumber \\
F_R&\to&F_R -  \left(\f Mm\right)^{2s} {m^2b_1^R\over 16\pi^{3/2}}
{\Gamma (s-1) \over \Gamma (s-{1\over 2})}\ ,\ 
\Lambda_R\to \Lambda_R - {mb_{3/2}^B\over 16\pi^{3/2}}\ ,\label{ren} \\
h_R&\to&h_R-\left(\f Mm\right)^{2s} {b_2^R\over 16\pi^{3/2}}
{\Gamma (s)\over\Gamma (s-{1\over 2})}\ . \nonumber
\en 
After this procedure we obtain the following expression for ground state 
energy
\be
E_q^R = -{m\over 32\pi^2 \beta}\left\{B^R(\beta)\ln\beta^2  
+ \Omega^R (\beta)\right\}\ . \label{EqR} 
\ee
The similar general structure of ground state energy in massless case has been 
obtained by Blau, Visser and Wipf \cite{ZetaFunction} using the dimensional 
considerations only. For massive case we find in manifest form the same 
structure. 
If we have used another scale for the mass like $M\to M/\chi$, in 
renormalization rules \Ref{ren}, the above logarithmic term $\ln\beta^2$ 
will be replaced by $\ln (\chi\beta)^2$. 

The expression \Ref{EqR} is, in fact, the Casimir energy for internal part of 
the spherical bag in global monopole background. For small radius of the bag, 
this energy tends to infinity as $\ln R/R$:
\be
E_q^R\sim -{m\over 16\pi^2 \beta}b_2^R\ln\beta\ ,  
\ee 
and for great radius of the bag $R\to\infty$ it tends to zero 
\be
E_q^R \sim -{mb_{5/2}^R\over 16\pi^{3/2} \beta^2}\ .
\ee
Using these two limits we may analyze qualitatively the dependence of the 
Casimir energy of the internal part of the bag on its radius. The behavior of 
energy is defined by two heat kernel coefficients $b_2^R$ and $b_{5/2}^R$. 
Both of these coefficients are the functions of non-minimal coupling 
parameter $\xi$ and metric coefficient $\alpha$. In general, three kinds of 
different behaviors exist, which are plotted in Fig.1. It is possible to 
analyze the energy in general case, however we shall discuss only three 
cases for $\xi = 1/6,\ 1/8,\ 0$. 

1. $\xi =\f 16$. In this case the behavior may be only of I and II 
kinds namely, the first kind for $\alpha < 1.24$ and the second one for 
$\alpha > 1.24$. The coefficient $b_2^R$ does not change its 
sign, but $b_{5/2}^R$ does for $\alpha = 1.24$. 

2. $\xi =\f 18$. The behavior of energy may be I, II and III kinds, the first 
kind is for $\alpha <1.045$, the second kind is in the region $1.045 <\alpha < 
1.17$ and the third one for $\alpha > 1.17$. In the point $\alpha = 1.045$ the 
coefficient $b_{5/2}^R$ change the sign, but $b_2^R$ does not up to $\alpha = 
1.17$ where it changes the sign, too. 

3. $\xi=0$. This case is similar to previously one: it is of the first kind 
for $\alpha < 1.016$, of the second for region $1.016 <\alpha <1.054$, and 
of the third for $\alpha >1.054$.       

For $\alpha \leq 1$ and $\xi=\f 16,\ \f 18,\ 0$ it is possible only the first 
kind of behavior. In the case when $\alpha = 1$ the energy was calculated 
numerically in Ref.\cite{Method3} and our results are in agreement with it.  
In this case $b_2^R=-\pi^2 16/315$ and $b_{5/2}^R=-\pi^{3/2}/120$ the 
dependence may be the of first kind, only. 

In the limit  $R\to\infty$, the quantum correction tends to zero and the full 
energy contains only classical part which are due to boundary and background 
itself. 

Let us now proceed  to our model. We surround the monopole origin by spheres 
of radii $r_0$ and $R>r_0$ and consider the bosonic matter field in the space 
between them. We do not take into account the interior of sphere of 
radius $r_0$, 
there is nothing inside it. We impose the Dirichlet boundary condition on this 
sphere which means that there is no flux into this region. The full energy in 
this case consists of five parts 
\be
E= E_{cl}^R + E_{cl}^{r_0} + E_{cl}^{gm} + E_q^R + E_q^{r_0}\ ,
\ee
where 
\bn
E_{cl}^R&=&p_RV_R +\sigma_R S_R + F_RR + \Lambda_R + \f {h_R}R\ , \\
E_{cl}^{r_0}&=&p_{r_0}V_{r_0} +\sigma_{r_0} S_{r_0} + F_{r_0}r_0 + 
\Lambda_{r_0} + \f {h_{r_0}}{r_0}\ ,\label{hr0} \\
E_{cl}^{gm}&=&-4\pi\eta^2\alpha (R-r_0)\ ,
\en
are the classical part of energy due to the boundaries and global monopole 
itself, respectively, and
\bn
E_q^R&=&\f 12 M^{2s}\zeta_A^R(s-{1\over 2})_{s\to 0}
= -{m\over 32\pi^2 \beta}\left\{B^R(\beta)\ln\beta^2  
+ \Omega^R (\beta)\right\}+\left(\f Mm\right)^{2s} {1\over 16\pi^{3/2}}
\nonumber\\ 
&\times&\left\{ B_0^R m^4
{\Gamma (s-2) \over \Gamma (s-{1\over 2})}
-\f 23 B_{\f 12}^R m^3  + B_1^R m^2 {\Gamma (s-1) \over \Gamma
(s-{1\over 2})} + B_{\f 32}^R m + B_2^R {\Gamma (s)\over\Gamma (s-{1\over 2})} 
\right\}_{s\to 0}\ ,\\ 
E_q^{r_0}& =& \f 12 M^{2s}\zeta_A^{r_0}(s-{1\over 2})_{s\to 0}
= -{m\over 32\pi^2 \beta_0}\left\{B^{r_0}(\beta_0)\ln\beta^2_0  
+ \Omega^{r_0} (\beta_0)\right\}+\left(\f Mm\right)^{2s} 
{1\over 16\pi^{3/2}}\nonumber\\ 
&\times&\left\{ B_0^{r_0} m^4
{\Gamma (s-2) \over \Gamma (s-{1\over 2})}
-\f 23 B_{\f 12}^{r_0} m^3  + B_1^{r_0} m^2 {\Gamma (s-1) \over \Gamma
(s-{1\over 2})} + B_{\f 32}^{r_0} m + B_2^{r_0} {\Gamma (s)\over\Gamma 
(s-{1\over 2})} \right\}_{s\to 0}\ ,
\en
are the quantum corrections. Adopting the same renormalization prescription 
given in Eq.\Ref{ren} for parameters in $E_{cl}^R$ and $E_{cl}^{r_0}$,  
one arrives 
at the following expression for renormalized quantum corrections
\be
E_q^R=-{m\over 32\pi^2 \beta}\left\{B^R(\beta)\ln\beta^2  
+ \Omega^R (\beta)\right\}\ \label{ER}
\ee
and
\be 
E_q^{r_0}=-{m\over 32\pi^2 \beta_0}\left\{B^{r_0}(\beta_0)\ln\beta^2_0  
+ \Omega^{r_0} (\beta_0)\right\}\ .\label{Er0} 
\ee
The sum of these terms gives the Casimir energy of the field in space between 
the two spheres in the global monopole background. The first part we have 
already discussed. We may consider the second part in the same way. For small 
radius of sphere, $r_0\to 0$, it tends to infinity 
\be
E_q^{r_0}\sim -{m\over 16\pi^2 \beta_0}b_2^{r_0}\ln\beta_0\ ,  
\ee 
and for great radius of sphere $r_0\to\infty$ it tends to zero 
\be
E_q^{r_0} \sim -{mb_{5/2}^{r_0}\over 16\pi^{3/2} \beta_0^2}\ .
\ee
Because $b_2^{r_0}=-b_2^R$ and $b_{5/2}^{r_0}=b_{5/2}^R$, the energy 
$E_q^{r_0}$ has different behavior at small radius $r_0$. The sum of $b_n^R$ 
and $b_n^{r_0}$ constitutes 
the whole heat kernel coefficients for this space. For this reason the three 
kinds of dependences of $E_q^{r_0}$ on the radius are possible, only, which 
are displayed in Fig.1. The same results are available for  $E_q^{r_0}$ as 
it was obtained above for $E_q^R$ for $\xi =1/6,\ 1/8,\ 0$ ; we have to 
change only the left 
plot to right one in Fig.1. The case $b_2=0$ has to be analyzed 
numerically, however this discussion is out the scope of present paper. 
  
For any non-zero radius of the inner cavity $r_0$ we have finite result. In 
this case the Casimir energy may be positive or negative, depends on the 
parameters of the theory. The main problem now is with the limit $r_0\to 0$,  
which has to reproduce the topological defect itself. The energy $E_q^{r_0}$ 
presents  divergence in this limit  as $\ln r_0/r_0$. 
This is in contradiction with above consideration of 
point-like monopole. If we set the  radius $r_0=0$ at the beginning as it 
was done in Sec.\ref{Zeta} we obtain zero ground state energy for 
$R\to\infty$. From the point of view of heat kernel coefficients we have 
already thrown away divergent part of $B_2$ using {\em partie finite} of 
integral \Ref{B2}.  In frameworks of our model this thrown part appears here 
as divergent at origin and the additional renormalization is needed. 

For renormalization we may use the last term $h_{r_0}/r_0$ in the classical 
part of energy which is due to boundary \Ref{hr0}. We define a parameter 
$M_0$ with dimension of mass by the relation $h_{r_0}=GM_0^2$, where $G$ is 
gravitational constant. With this definition, the divergent contribution for 
small radius $r_0$ may be canceled by the renormalization rule below
\be
M_0^2\to M_0^2 + \f{m_{pl}^2}{32\pi^2}\left\{2B^{r_0}\ln(mr_0) +
\Omega^{r_0} (mr_0)\right\}\ ,\label{Ren}
\ee
where $m_{pl}^2$ is the Plank mass, and then ground state energy is zero.  

\section{Conclusion}\label{last}

In this paper we have considered the ground state energy of quantum scalar 
field in the background of global monopole space-time, with line element given 
by Eq.\Ref{Metrica}, in framework of zeta function approach. In order to 
reveal the role of singularity, we investigated two cases: In the first case 
we calculated ground state energy for point-like global monopole. We 
surrounded the origin by sphere of  radius $R$ and obtained that the ground 
state energy of the field inside. It has the form \Ref{EqR} and tend to zero 
in the limit $R\to \infty$ and to infinity when $R\to 0$. The behavior of this 
Casimir energy in this cases is managed by two heat 
kernel coefficients $b_2^R$ and $b_{5/2}^R$, respectively. The qualitative 
plots of the ground state energy for different values of the parameters 
$\xi =1/6,\ 1/8,\ 0$ and $\alpha$ are given in Fig.1(a). For $\alpha \leq 1$ 
it may be only in first kind for above values of $\xi$. 

In order to avoid the problem with singularity at origin, we investigated the 
second case in Sec.\ref{Zeta1}, the following model: We surrounded the origin 
by a sphere of 
radius $r_0<R$ and considered the scalar field in domain between these two 
spheres using the  Dirichlet boundary condition on the wave function 
associated with the massive scalar field, on the two surfaces. This boundary 
condition guarantee that there is no flux of particle through the spherical 
surfaces. The Casimir energy in this case consists of two parts given in 
Eqs.\Ref{ER}, \Ref{Er0}. The first part is the same as for pint-like monopole 
case  and second one is due to the inner sphere of radius $r_0$ around 
origin. The structure of second part of ground state energy \Ref{Er0} is 
similar: there is logarithmic divergence at origin which tends to zero for 
infinite radius. The sign of energy for 
small distance is opposite. For this reason there appears three kind of 
dependence of energy which are displayed in Fig.1(b).   

In the limit $R\to\infty$ and finite $r_0\not= 0$ only one contribution in 
ground state energy  survives \Ref{ER}. In the limit $r_0\to 0$, it is 
divergent and additional renormalization is needed which is given by 
Eq.\Ref{Ren}. After this renormalization the ground state energy of a global 
monopole will be zero. This is in agreement with ground state energy of 
point-like global monopole.  

If one fills up this cavity around the origin by matter, the situation 
becomes different. We may expect the same kind of divergence for additional 
energy from the interior of monopole but with opposite sign. 
We have already seen that the internal and external contributions have 
opposite signs: $b_2^R=-b_2^{r_0}$. For this reason we may expect that 
this kind of divergence will cancel, however we cannot say anything 
analytically about divergence $\Omega (mr_0)/r_0$. In flat space-time 
\cite{Method3} it cancels, too, because the ground state energy is 
zero for zero radius of the bag. The same cancellation takes place in the 
case of a thick cosmic string background considered in 
Ref.\cite{KhusnutdinovBordag}. All of these aspects will be discussed in a 
separate paper.   

\section*{Acknowledgments}
NK would like to thank Dr. M. Bordag for many helpful
discussions, NK is also grateful to Departamento de
F\'{\i}sica, Universidade Federal da Paraiba (Brazil) where this
work was done, for hospitality. His work was supported in part
by CAPES and in part by the Russian Found for Basic Research,
grant No 99-02-17941.  

ERBM and VBB also would like to thank the Conselho Nacional de
Desenvolvimento Cientifico e Tecnol\'ogico (CNPq). 

\appendix
\section{} \label{appA}

In this appendix we want to give some brief explanation about the most 
important results found by us. 
First of all we represent the expressions for $A_k$ as series in powers of
 $\beta = m r_0$ regarding for a moment $b,\beta < 1$ and $\Delta > -1/4$ for 
convergence of series. They are 
\bn
A_{-1}(s,R)&=& \f m\beta \beta^{2s}{4\pi \over \alpha} \sum_{n=0}^\infty 
{(-1)^n \over n!}\beta^{2n} {\Gamma (s-1+n) \over s+n-\f 12} \sum_{l=0}^\infty 
{\nu_l \over (\nu_l^2 + \Delta)^{s+n-1}}\ , \\
A_0(s,R)&=& -\f m\beta \beta^{2s} 4\pi^{3/2}\sum_{n=0}^\infty {(-1)^n 
\over n!} b^{2n} \Gamma (s + n - \f 12) \zeta_H (2s+2n-2, \f 12)\ , \\ 
A_k(s,R)&=& -\f m\beta \beta^{2s}16\pi^{3/2}\alpha^k \sum_{n=0}^\infty {(-1)^n 
\over n!}b^{2n} \sum_{a=0}^k x_{k,a} {\Gamma (s+\f k2+a+n - \f 12) \over 
\Gamma (\f k2 + a)}\nonumber\\ 
&\times&\sum_{l=0}^\infty {(\nu_l^2 + \Delta)^a \over \nu_l^{2s + 
k + 2a +2n -2}} \ . 
\en 
In the above formulas we have used the following notations : 
$\nu_l = l + \f 12\ ,\ \Delta = 2(1 - \alpha^2)(\xi - 1/8)$ and 
$b^2 = \beta^2 + \Delta$. As we can see only the first three terms in the 
expression for  $A_{-1}$ , with $n=0,1,2$, two terms in $A_0\ ,\ A_1\ ,\ A_2$ 
and one term in $A_3$ are divergent in the limit $s \to 0$. Extracting these 
terms we may set $s=0$ in the remaining series and we get the following 
result: 
\bn
\zeta_A^R(s-\f 12) &=& N^R(s,\beta) + {m^{-2s}\over (4\pi)^{3/2}\Gamma (s-
\f 12)}\sum_{k=-1}^3 A_k(s,R)\nonumber \\
&=& {m^{-2s}\over (4\pi)^{3/2}\Gamma (s-
\f 12)} \beta^{2s}\left\{m^4 B_0^R \Gamma (s-2) +m^3 B_{\f 12}^R 
\Gamma (s-\f 32)+ m^2 B_1^R \Gamma (s-1)\nonumber\right.\\ 
&+&\left.m B_{\f 32}^R \Gamma (s-\f 12)+  B_2^R \Gamma (s)\right\}
-{1\over 16 \pi^2 R}\left\{ \sum_{k=-1}^3 \omega_k(\beta) + 
\omega_f^R(\beta)\right\}, 
\en    
where 
\bn
\omega_f^R(\beta)&=&32\pi \sum_{l=0}^\infty \nu_l \sqrt{\nu_l^2 + \Delta}
\int_{\beta/\sqrt{\nu_l^2 + \Delta}}^\infty dx \sqrt{x^2 - {\beta^2 \over 
\nu_l^2 + \Delta}}\nonumber\\ 
&\times&{\partial \over \partial x} \left\{ \ln I_\mu(\mu x) - 
\mu\eta (x) + \f 14 \ln (1 + x^2) - \f 1\mu D_1  - \f 1{\mu^2} D_2 - 
\f 1{\mu^3} D_3\right\}\ ,\\ 
\omega_{-1}(\beta )&=&-{4\pi \over\alpha}\left\{\left[ -\f 72 \zeta'(-3) - 
\f 7{160}+\f 1{240}\ln 2+\Delta \left( - 2\zeta'(-1) + \f 16 - 
\f 16\ln2\right)\right]\right.\nonumber\\ 
&+&\beta^2 \left[ 2\zeta'(-1)+ \f 14 + \f 16\ln2 + \Delta \left( 
2\gamma + 4\ln 2 -3\right)\right] + \beta^4 \left[\f 13\gamma + \f 23\ln 2 - 
\f {13}{36}  \right]\nonumber\\
&-&\left.\sum_{n=3}^\infty {(-1)^n \over n!}\Gamma (n-1) \Delta^n 
\zeta_H (2n-3, \f 12)  -  2\beta^2\sum_{n=2}^\infty{(-1)^n \over n!}
\Delta^{n}\Gamma (n) \zeta (2n-1)\right.\nonumber\\
&+&\left.\f 13 \beta^4 Y_{1,1}(\Delta )
+\sum_{n=3}^\infty {(-1)^n \over n!}\f{\Gamma (n-1)}{n-\f 12}
\beta^{2n} \sum_{l=0}^\infty \f{\nu_l}{(\nu_l^2 + \Delta)^{n-1}}\right\}\ ,\\
\omega_0(\beta)&=& -2\pi^2 \sum_{n=2}^\infty{(-1)^n \over n!}
{\Gamma (n-\f 12)\over \Gamma (\f 32)} b^{2n} \zeta_H (2n-2, \f 12)\ , \\ 
\omega_1(\beta)&=& -2\pi\alpha \left\{\left[-\zeta'(-1) - \f 5{36} - 
\f 1{12}\ln 2 + \Delta \left(-\gamma - 2\ln 2 + \f 53 \right)\right]\right.
\nonumber\\ 
&+&\beta^2 \left[\f 73\gamma + \f 12 + \f {14}3 \ln2\right] 
+\sum_{n=2}^\infty {(-1)^n \over n!}\left(\Gamma (n) - \f{10}3 
\Gamma (n+1)\right) b^{2n} \zeta_H (2n-1, \f 12)\nonumber\\ 
&-&\left. \f{10}3\Delta Y_{1,1}(b)\right\}\ , \\
\omega_2(\beta)&=& -2\pi^2\alpha^2 \left\{\f 3{16}\pi^2\Delta + 
\f 5{32}\pi^2 \Delta^2 + \f 12 Y_{\f 12,0}(b) -  \f 32 Y_{\f 32,0}(b) 
+  \f{15}{16} Y_{\f 52,0}(b)\right.\nonumber \\
&+&\left. \Delta \left[ - \f 32 Y_{\f 32,2}(b)+\f{15}8 Y_{\f 52,2}(b)\right]
+\Delta^2  \f{15}{16} Y_{\f 52,4}(b)\right\}\ ,\\
\omega_3(\beta )&=& -2\pi\alpha^3\left\{\f{293}{1512} - \f{22}{2520}\gamma 
- \f{229}{1260}\ln 2  + \left[\f{25}{24} X_{1,1}(b) - \f{177}{20} X_{2,1}(b)
\right.\right.\nonumber\\
&+&\left.\f{221}{15} X_{3,1}(b)-\f{442}{63} X_{4,1}(b)\right]+
\Delta \left[-\f{177}{20} Y_{2,3}(b) + \f{442}{15} Y_{3,3}(b) - \f{442}{21} 
Y_{4,3}(b) \right]\nonumber\\ 
&+&\left. \Delta^2 \left[\f{221}{15} Y_{3,5}(b) -\f{442}{21} Y_{4,5}(b)\right]
-\Delta^3 \f{442}{63} Y_{4,7}(b)\right\}\ , \\
X_{p,q}(b)&=&\sum_{n=0}^\infty{(-1)^n \over n!}{\Gamma (n+p) \over \Gamma (p)} 
b^{2n}\zeta_H(2n + q,\f 12)\ ,\\
Y_{p,q}(b)&=&\sum_{n=1}^\infty{(-1)^n \over n!}{\Gamma (n+p) \over \Gamma (p)} 
b^{2n}\zeta_H(2n + q,\f 12)\ .
\en
Each of the above series may be analytically continued in terms of digamma 
function $\Psi$ for arbitrary values of $b$ and $\Delta$. For example 
\bn
&&\sum_{n=3}^\infty {(-1)^n \over n!}\f{\Gamma (n-1)}{n-\f 12}
\beta^{2n} \sum_{l=0}^\infty \f{\nu_l}{(\nu_l^2 + \Delta)^{n-1}} = -2\beta^4 
\int_0^1 dx x (1-x)^2 \\
&\times&\left\{ \Psi\left[\f 12 -i\sqrt{\Delta +\beta^2 x^2}\right] + 
 \Psi\left[\f 12 +i\sqrt{\Delta + \beta^2 x^2}\right]- \Psi\left[\f 12 -
i\sqrt{\Delta}\right] -  \Psi\left[\f 12 +i\sqrt{\Delta}\right] \right\}\ , 
\nonumber \\
&&\sum_{n=3}^\infty {(-1)^n \over n!}\Gamma (n-1) \Delta^n 
\zeta_H (2n-3, \f 12)\nonumber\\ 
&=& -\Delta^2 \int_0^1 dx x (1-x^2)
\left\{\Psi\left[\f 12 - i\sqrt{\Delta}\right] + 
\Psi\left[\f 12 +i\sqrt{\Delta}
\right] - 2\Psi\left[\f 12\right]\right]\ . 
\en
This kind of representation is suitable for numerical calculations.

Adding and subtracting the asymptotic expansion of zeta function \Ref{ze} we 
obtain the following formula 
\bn
\zeta_A^R(s-\f 12)&=& -{m\over 16\pi^2 \beta}\left\{B^R(\beta)\ln\beta^2  
+ \Omega^B (\beta)\right\}+{m^{-2s} \over (4\pi )^{3/2}}\left\{ B_0^R m^4
{\Gamma (s-2) \over \Gamma (s-{1\over 2})}\right.\nonumber\\
&+&\left. B_{\f 12}^R m^3 {\Gamma (s-{3\over 2})
\over \Gamma (s-{1\over 2})} + B_1^R m^2 {\Gamma (s-1) \over \Gamma
(s-{1\over 2})} + B_{\f 32}^R m + B_2^R {\Gamma (s)\over\Gamma (s-{1\over 2})} 
\right\}\ , 
\en
where $\Omega^B (\beta )  = \sum_{k=-1}^3 \omega_k(\beta) + 
\omega_f^R(\beta)$. It is easy to see that the function $\Omega^B (\beta )$ 
tends to a constant which may be calculated using the above 
formulas. Indeed, in the limit $\beta\to \infty$ we have to obtain asymptotic 
expansion of zeta function. Because one has already extracted the first five 
terms, so the next term will be $B_{5/2}$. For this reason we get the 
following 
behavior for great $\beta$ : $\Omega^B (\beta ) \sim -B^R \ln\beta^2 + 
\sqrt{\pi}b_{\f 52}^R/\beta +...$ . The coefficient $b_{\f 52}^R$ may be 
found in the same way and it has the following form 
\be
b_{\f 52}^R = \f Rm B_{\f 52}^R = \pi^{3/2}\left\{ \f 1{16} 
\alpha^4 + \f 12 \left(\Delta - \f 1{12}\right)\alpha^2 - \left(\Delta^2 - 
\f 16 \Delta + \f 7{240}\right)\right\}\ .
\ee   

It is easy to obtain the formulas for zeta function $\zeta_A^{r_0}$ from above 
expressions. The index $k$ in $A_k$ corresponds to a term which is 
proportional 
to $\mu^{-k}$ in uniform expansion of the Bessel function in Eq.\Ref{unex}. 
The uniform expansion of the modified Bessel function of second kind given in 
Eq.\Ref{UniformExpansionofK} may be obtained from uniform expansion of the
modified Bessel function of first kind \Ref{UniformExpansion} by replacing   
the index $\mu$ to $-\mu$. For this reason in order to obtain the formulas 
for  the zeta function $\zeta_A^{r_0}$, we may replace $R\to r_0,\ \beta\to 
\beta_0$ and $A_k(s,R) \to (-1)^kA_k(s,r_0)$ in the above formulas for 
zeta function $\zeta_A^R$. As the odd degrees of $\mu$ give contributions to  
heat kernel coefficients with integer index, they will change the sign. 
Therefore we have the following formula for the  zeta function $\zeta_A^{r_0}$ 
\bn
\zeta_A^{r_0}(s-\f 12)&=& -{m\over 16\pi^2 \beta_0}\left\{B^{r_0}(\beta_0)
\ln\beta_0^2  + \Omega^{r_0} (\beta_0)\right\}+ 
{m^{-2s} \over (4\pi )^{3/2}} \left\{ B_0^{r_0} m^4
{\Gamma (s-2) \over \Gamma (s-\f 12)}\right.\nonumber\\
&+&\left.B_{\f 12}^{r_0} m^3 {\Gamma (s-\f 32)
\over \Gamma (s-\f 12)} + B_1^{r_0} m^2 {\Gamma (s-1) \over \Gamma
(s-\f 12)} + B_{\f 32}^{r_0} m + B_2^{r_0} {\Gamma (s) \over \Gamma (s-\f 12)} 
\right\}\ , 
\en   
where $\Omega^{r_0} (\beta_0 )  = \sum_{k=-1}^3(-1)^k \omega_k(\beta_0) + 
\omega_f^{r_0}(\beta_0)$ and 
\bn
\omega_f^R(\beta)&=&32\pi \sum_{l=0}^\infty \nu_l \sqrt{\nu_l^2 + \Delta}
\int_{\beta/\sqrt{\nu_l^2 + \Delta}}^\infty dx \sqrt{x^2 - {\beta^2 \over 
\nu_l^2 + \Delta}}\\ 
&\times&{\partial \over \partial x} \left\{ \ln K_\mu(\mu x) + 
\mu\eta (x) + \f 14 \ln (1 + x^2) + \f 1\mu D_1  - \f 1{\mu^2} D_2 + 
\f 1{\mu^3} D_3\right\}\ . \nonumber 
\en 


\newpage
\centerline{Figure captions}
Figure 1. Three types of dependence of the ground state energy for the field 
inside the sphere of radius $R$ (a) and outside the sphere of radius 
$r_0$ (b).
\newpage 
\centerline{Figures}

\vspace*{5cm}
  \centerline{
\psfig{figure=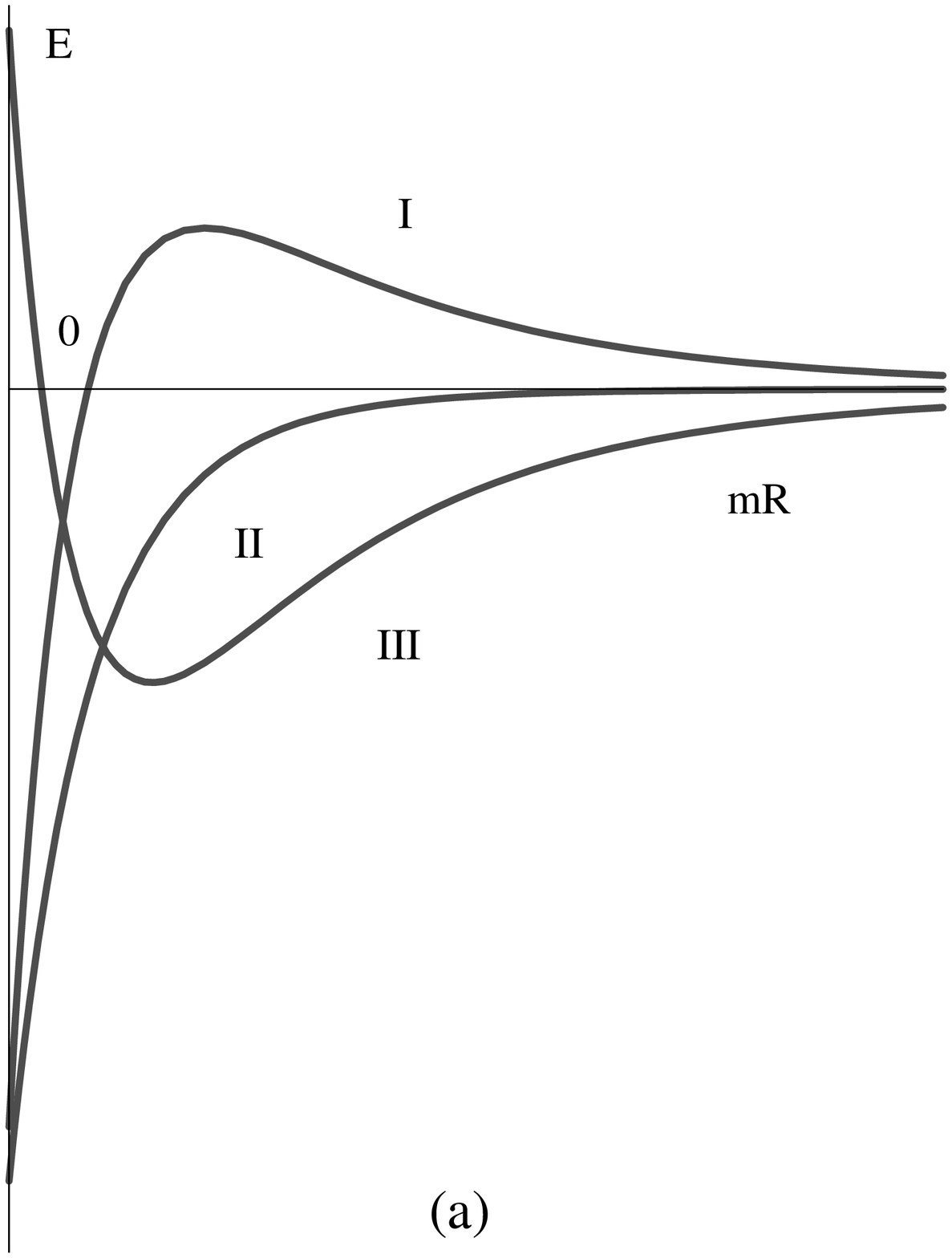,angle=0,height=10cm}%
\psfig{figure=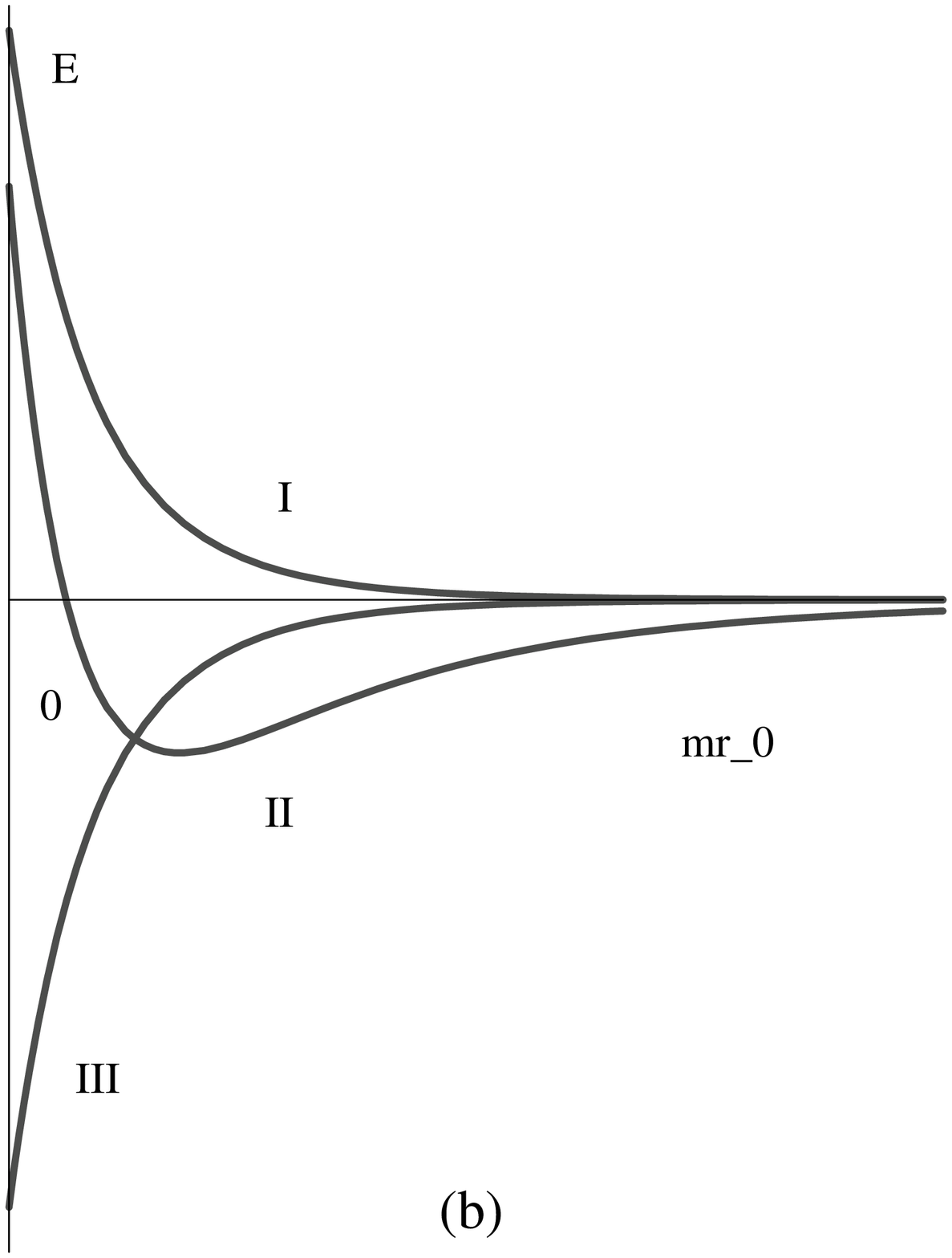,angle=0,height=10cm}}
\newpage
\centerline{Footnotes}
\begin{description}
\item[*] e-mail:emello@fisica.ufpb.br 
\item[\dag] e-mail: valdir@fisica.ufpb.br
\item[\ddag] On leave from Kazan State Pedagogical University, Kazan, 
Russia; e-mail:nail@dtp.ksu.ras.ru
\end{description}
\end{document}